\documentstyle[prl,aps,preprint]{revtex}
\begin{document}
\draft
\title{Superconductivity in the three-leg Hubbard ladder:
a Quantum Monte Carlo study}
\author{Takashi Kimura, Kazuhiko Kuroki, and Hideo Aoki}
\address{Department of Physics, University of Tokyo, 
Hongo, Tokyo 113, Japan}
\date{\today}
\maketitle
\begin{abstract}
Quantum Monte Carlo method is used to look 
into the superconductivity 
in the three-leg Hubbard ladder. 
The enhanced correlation for the pairing 
across the central and edge 
chains, which has been predicted in the weak-coupling 
renormalization as an effect of coexistence of 
gapful and gapless spin modes, 
is here shown to persist for intermediate 
interaction strengths. 
\end{abstract} 

There is an increasing fascination toward 
strongly correlated electrons on ladders. 
This has been kicked
off by the pioneering works \cite{Schulz,Rice}, 
which have proposed that the ladder 
at half-filling should be a spin liquid 
with gapful spin excitations for an even number of chains, 
while odd-numbered chains should be antiferromagnetic (AF) with 
gapless spin excitations. 
This property, which is reminiscent of Haldane's conjecture 
\cite{Schulz,Haldane,Affleck} for the one 
dimensional (1D) AF Heisenberg model for integer 
and half-odd-integer spins, have been confirmed 
by numerical calculations\cite{Dagotto,Greven}.  
Experimentally the ladders realized in cuprates 
indeed exhibit a spin liquid behavior 
when two-legged, while a three-leg system 
shows an AF behavior. \cite{Ishida}

This has led to an 
expectation that a doping of the ladder with carriers 
will produce 
an interchain singlet superconductivity 
in an even-numbered ladder 
associated with the persistent spin gap, 
while an odd-numbered ladder should have the usual 
$2k_F$ spin-density
wave (SDW) reflecting the gapless spin excitations.
\cite{Rice,Dagotto2}  
The superconductivity in two-leg ladders has indeed been 
confirmed by the perturbational
renormalization group for weak-coupling repulsive 
interactions.\cite{Fabrizio,Finkel,Schulz2,Balents,Nagaosa}
Most of the numerical calculations for $t-J$ ladder models 
also support the dominant pairing 
correlation.\cite{Dagotto2,Tsunetsugu,Hayward,Sano} 
The numerical results for the Hubbard ladder 
models\cite{Yamaji,Noack,Asai} 
are less conclusive, but some of the controversies 
have recently been resolved from Quantum Monte Carlo 
(QMC) calculations, 
where the pairing correlation is indeed enhanced 
in Hubbard ladders.\cite{Kuroki}  
Experimentally a occurrence of superconductivity is 
reported for a cuprate recently\cite{Uehara}.  

Now, whether the `even-odd' conjecture 
for superconductivity continues 
to be valid for triple chains has to be tested.  
In fact, we \cite{Takashi} and Schulz \cite{Schulz3} 
have independently calculated the
correlation functions in the three-leg Hubbard ladder 
within the weak-coupling theory and
have found similar results. 
The message obtained there is that 
an odd number (three) of legs can in fact superconduct, 
providing a counter-example of the even-odd conjecture.   
A key is that 
gapless and gapful spin excitations coexist in a three-leg ladder. 
This has been analytically shown from the correlation functions 
starting from the phase diagram given by Arrigoni\cite{Arri}, 
who has enumerated the number of gapless and gapfull modes 
with the perturbative renormalization-group technique
in the weak-coupling limit.  
The coexisting gapful and gapless modes gives rise to 
a peculiar situation where a 
specific pairing {\it across} the central and 
edge chains (that is roughly a d-wave pairing) is dominant, 
while the $2k_F$ SDW on the edge chains 
simultaneously shows a subdominant but still 
long-tailed (power-law) decay 
associated with the gapless spin mode.\cite{comment} 
In other words, the dominant 
superconductivity only requires the existence
of gap(s) in not all but {\it some} of the spin modes 
when there are multiple of them.

However, there is a serious question about 
these weak-coupling approaches.  
First, only for an infinitesimally small coupling is the 
weak-coupling theory guaranteed to be valid in principle.
Furthermore, when there is a gap in the excitation, 
the renormalization flows into a strong-coupling regime, 
so that the weak-coupling (perturbational)
theory might break down even for small $U$.  
Hence it is imperative to study the problem from an independent 
numerical method, especially 
for an intermediate strength of the Hubbard $U\sim t$. 
Although such a study comparing the numerical result for $U\sim t$ 
with the weak-coupling theory has been done for 
the two-leg system, this does not necessarily serve 
to enlighten 
the situation in the three-leg case, where
gapless and gapful modes coexist. This is exactly our motivation 
for the present study, which reports an extensive QMC calculation 
for three-leg Hubbard ladder.  
The result indeed turns out to exhibit 
an enhancement of the pairing correlation even 
for finite coupling constant, $U/t=1\sim 2$.

The Hamiltonian of the three-leg Hubbard model 
is given by in standard notations as
\begin{eqnarray}
H=&&-t\sum_{\mu i\sigma}
(c^{\dagger}_{\mu i \sigma}c_{\mu i+1 \sigma}+{\rm h.c.})
\nonumber\\
&&-t_{\perp}\sum_{i\sigma}
(c^{\dagger}_{\alpha i\sigma}c_{\beta i\sigma}+
c^{\dagger}_{\beta i\sigma}c_{\gamma i\sigma}+{\rm h.c.})
\nonumber\\
&&+U\sum_{\mu i}n_{\mu i\uparrow}n_{\mu i\downarrow}, 
\end{eqnarray}
where $t (t_{\perp})$ is the intra-(inter-)chain hopping, 
$i$ labels the rung while $\mu = \alpha , \beta , \gamma$ labels the 
leg (with $\beta$ being the central one).  
In the momentum space we have
\begin{eqnarray}
H=&& \sum_{k\sigma}\left( -2t{\rm cos}(k)-\sqrt{2}t_{\perp}\right)
a^{\dagger}_{1k\sigma} a_{1k\sigma}\nonumber\\
&&-2t\sum_{k\sigma} {\rm cos}(k) a^{\dagger}_{2k\sigma} a_{2k\sigma}
\nonumber\\
&&+\sum_{k\sigma} \left( -2t{\rm cos}(k)+\sqrt{2}t_{\perp}\right)
a^{\dagger}_{3k\sigma}a_{3k\sigma}
\nonumber\\
&&+U\sum({\rm terms\  of\  the\  form}\ a^{\dagger}a^{\dagger}aa). 
\end{eqnarray}
Here $a_{jk\sigma}$ annihilates 
an electron with lattice momentum $k$ in the $j$-th band
($j=1,2,3$), where $a_{jk\sigma}$ is related to 
$c_{\mu k\sigma}$ (the Fourier transform of $c_{\mu i\sigma}$) 
through a linear transformation,
\begin{eqnarray}
\left( \begin{array}{c} 
c_{\alpha k\sigma} \\ c_{\beta k\sigma} \\ c_{\gamma k\sigma}
\end{array} \right)
=
\left( \begin{array}{ccc}
\frac{1}{2} & \frac{1}{\sqrt{2}} & \frac{1}{2}\\
\frac{1}{\sqrt{2}} & 0 & -\frac{1}{\sqrt{2}}\\
\frac{1}{2} & -\frac{1}{\sqrt{2}} & \frac{1}{2}
\end{array} \right)
\left( \begin{array}{c}
a_{1k\sigma} \\ a_{2k\sigma} \\ a_{3k\sigma}
\end{array} \right).
\end{eqnarray}
 
If we first recapitulate the weak-coupling theory
\cite{Takashi,Schulz3,Arri}, 
the pair hopping process 
($a^{\dagger}_{1\uparrow}a^{\dagger}_{1\downarrow}
a_{3\downarrow}a_{3\uparrow}+$h.c.) between
the first and third bands and the backward-scattering 
processes within the first or third band become relevant
scattering processes. As a result,
two spin modes and one charge mode become gapful. 
This leaves one spin mode and two charge modes gapless, 
which are characterized by critical exponents, 
$K^*_{\sigma2}(=1$ 
for the spin independent interaction) and 
$K^*_{\rho2}$, $K^*_{\rho3}$, 
respectively\cite{Takashi}.  
We can recognize that the first (third) band 
is analogous to the bonding (anti-bonding) band in
the two-leg ladder, while the second band is analogous 
to the single 1D (Luttinger-liquid like) system. 

The correlation of the intraband 
singlet pairing within the first or third band, 
$\sum_{\sigma}\sigma(a_{1k\sigma}a_{1k-\sigma}
-a_{3k\sigma}a_{3k-\sigma})$, decays like 
$r^{-(1/K^*_{\rho2}+1/2K^*_{\rho3})/3}$ 
at large distances. 
This pair, when expressed in a real space via the 
inverse Fourier transform, is an interchain 
singlet pair across the central chain and an 
edge chain, 
$O_i= (c_{\alpha i\sigma}+c_{\gamma i\sigma})c_{\beta i-\sigma}- 
(c_{\alpha i-\sigma}+c_{\gamma i-\sigma})c_{\beta i\sigma}$. 
All the $K^*_{\rho}$'s should tend to unity in the limit of 
vanishing interaction, where 
the interchain superconductivity has an exponent of 1/2 while 
 the density wave correlations 
have exponents of at least 2.\cite{Takashi,Schulz3}
Thus the pairing correlation is identified as the most dominant.  
In the weak-coupling renormalization, however, 
we have to make a reasoning: 
`the pair hopping process and the backward scattering process 
flow, in the weak-coupling renormalization, 
into the strong-coupling regime upon our 
integrating out the high-energy modes, which 
results in a formation of the gaps'. 
Thus the validity of the weak-coupling scheme has to be tested 
as stressed above.  
This problem should be especially subtle when 
gapful and gapless modes coexist.

Here we employ the projector Monte Carlo method\cite{Monte}
to look into the ground state
pairing correlation function 
$P(r)\equiv\langle O^{\dagger}_jO_{j+r}\rangle$.  
We assume periodic boundary conditions
along the chain direction, $c_{N+1}\equiv c_1$, where $N$
is the number of rungs. We only consider here the case where the 
intra- and inter-band Umklapp processes are irrelevant because
that is the case where the above mentioned result 
obtained by weak-coupling 
theory is valid. 
The details of the QMC calculation are similar
to those for our calculation for the two-leg case.\cite{Kuroki}  
Specifically, the negative sign problem makes the QMC calculation 
feasible for $U\leq 2t$.  
We set $t=1$ hereafter.

In the two-leg case with a finite $U$, 
we have found an interesting property for 
finite systems: 
the pairing correlation is enhanced 
in agreement with the weak-coupling theory 
only when the one-electron energy
levels of the bonding and anti-bonding bands lie close to 
each other around the Fermi level 
(which is certainly the case with an infinite system).\cite{Kuroki} 
When the levels are misaligned 
(for which a 5\% change in $t_{\perp}$ is enough), 
the enhancement of the pairing correlation dramatically vanishes.  
In the weak-coupling theory, ratio of 
the spin gap to the level offset is assumed to be infinitely large 
at the fixed point of the renormalization flow, 
so that the level offset has to be small for 
the effect of the spin gap to be detectable in a finite system.

We have found that this applies to the three-leg ladder 
as well, i.e., 
the pairing correlation is enhanced 
when the one-electron levels of 
the first and third bands lie close to each other. 
Hence we concentrate on such cases hereafter.

We first show in Fig.1 the result for $P(r)$ for $t_{\perp}=0.92$ 
with $U=1$ with the band filling
$n=0.843=86$ electrons/(34 rungs$\times$ 3 sites). 
For this choice of $t_{\perp}$ the levels in 
first and third bands lie close to each
other around the Fermi level with the level 
offset being as small as 0.01.  
We can see that there exists 
a large enhancement over the $U = 0$ result at large distances. 
This is the key result of this paper.

Although it is difficult
to determine the decay exponent of $P(r)$, 
we can fit the data by supposing a trial function
as expected from the weak-coupling theory 
as we did in the two-leg case 
\cite{Kuroki},
\begin{eqnarray}
P(r)=&&\frac{1}{\pi^2}\sum_{d=\pm}
\{ cr^{-1/2}_d + (2-c)r^{-2}_d\nonumber\\
&&-[{\rm cos}(2k_{F1}r_d)+{\rm cos}(2k_{F3}r_d)]r^{-2}_d\}.
\end{eqnarray}

Here $k_{F1}(k_{F3})$ is the non-interacting Fermi wave number
of the first (third) band, while a constant $c$, 
which should vanish for $U=0$, 
is here least-square fit (by taking logarithm of the data)
as $c = 0.05$. 
Since we assume the periodic boundary condition, 
we have to consider contributions from both ways around, 
so there are two distances between the 0-th and 
the r-th rung, i.e.,
$r_+ = r$ and $r_- = N - r$. 
The overall decay should be $1/r^2$ 
as in the single-chain case, while the term 
$c/r^{1/2}$, the dominant 
correlation at large distances, is borrowed from 
the weak-coupling result.\cite{Takashi,Schulz3} 
The QMC result for a finite $U = 1$ 
fits to the trial form (solid line in Fig.1) 
surprisingly accurately.  
A finite $U$ may give some corrections to these functional forms, 
but even when we best-fit the exponent itself as $c/r^{\alpha}$ 
in place of $c/r^{1/2}$, 
we obtain $\alpha <0.7$ with a similar accuracy.

In Fig.2, we show the result for a larger
interaction $U = 2$. 
The result again shows the enhanced pairing correlation at large
distances.
However, the enhancement is slightly reduced 
than the $U = 1$ case. 
This is consistent with the weak-coupling theory, 
in which $K^*_{\rho}$'s should be a decreasing function of $U$.

Finally, we study if the presence of band 2 around 
$E_F$ can be detrimental to superconductivity.  
In Fig.3, 
we make the one-electron energy levels of all the three bands 
lie close to each other around
the Fermi level. 
This is accomplished here 
for $t_{\perp}= 0.685$ 
and the band filling $n = 0.719 = 82$ electrons/(38 rungs $\times$
3 sites). 
The highest occupied level of the second band then lies 
between that of the first band 
and the lowest unoccupied level of the
third band (lying above the highest 
occupied level of the first band by as small as 0.01, 
inset of Fig.3).  

The result in Fig.3 for $U = 1$ shows that 
the pairing correlation is enhanced as well. Thus we may
consider that the second band does not hinder the superconductivity 
in other bands.  
This is also consistent with the weak-coupling theory, 
in which all of the scattering processes connected with 
the second band are irrelevant.   
The fit of the correlation function to the trial one is again 
excellent with $c = 0.03$.

It is intriguing to investigate how the 
intermediate-$U$ regime (which is 
shown in the present study to be similar to the weak-coupling 
situation) would cross over to the large-$U$ Hubbard model.
It is also important to 
study the pairing correlation function in the three-leg $t-J$ ladder 
in order to clarify similarities and differences between $t-J$
and Hubbard ladders, since the former with an infinitesimal 
$J\sim t^2/U$ is an effective Hamiltonian of the latter with 
large $U$.      
The crossover to larger (especially odd) numbers of legs 
in the Hubbard model is also of interest.

We are grateful to Professor H.J. Schulz 
for attracting our attention to Ref.\cite{Schulz}and\cite{Schulz3}.  
Numerical calculations were done on 
FACOM VPP 500/40 at the Supercomputer Center,
Institute for Solid State Physics, 
University of Tokyo and on HITAC S3800/280 at the
Computer Center of the University of Tokyo.  
For the latter facility we thank Prof.  Y.
Kanada for a support in `Project for Vectorized Super Computing'.  
This work was also
supported in part by Grand-in-Aid 
for Scientific Research from the Ministry of Education of Japan. 
One of the authors (T.K.) acknowledges the Japan
Society for the Promotion of Science for a fellowship.

\begin{figure}
\caption{
The QMC result for the pairing correlation function, 
$P (r)(\bullet)$, plotted against the
real space distance $r$ in a three-leg Hubbard ladder 
with 34-rung having 86 electrons for $U = 1$
with $t_{\perp} = 0.92$. 
The dashed line is the non-interacting result 
for the same system size, while the
straight dashed line represents $\sim r^{-2}$. 
The solid line is a fit to a trial function (see text).
}
\end{figure}
\begin{figure}
\caption{
A similar plot as in Fig.1 for $U = 2$.
}
\end{figure}
\begin{figure}
\caption{
A similar plot as in Fig.1 for a 38-rung system 
having 82 electrons for $U = 1$ with $t_{\perp} = 0.685$.
The inset schematically depicts the positions of 
energy levels for the non-interacting case.}
\end{figure}

\begin{references}
\bibitem{comment} The power-law decay of the 
2$k_F$ SDW in doped systems suggests that 
the SDW correlation decays even more slowly at half-filling 
where Umklapp processes should
enhance the SDW. 
This is consistent with the NMR experiment\cite{Ishida}, 
in which an AF correlation is found.
\end{references}
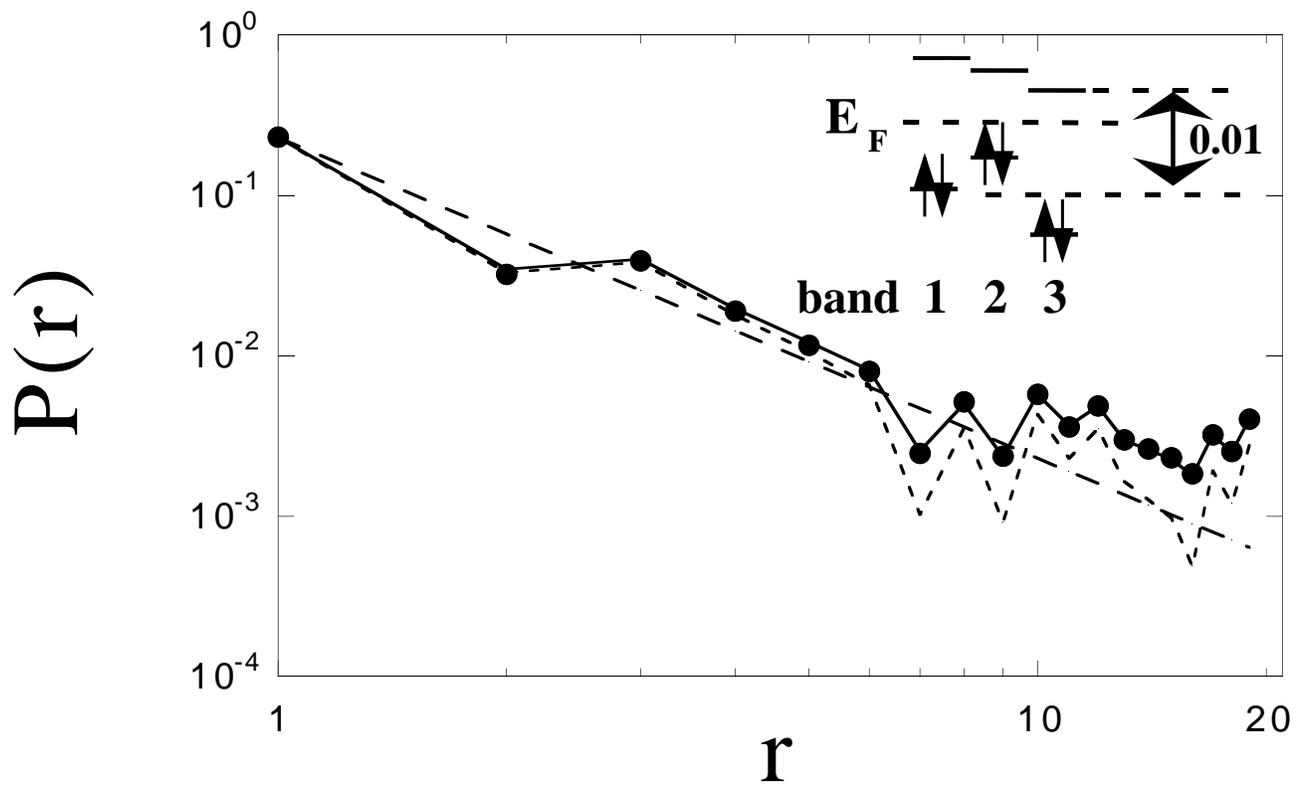
\end{document}